\documentstyle[11pt,aaspp4,flushrt,tighten]{article} 
\input{psfig.sty}
\newcommand{\ltsima} {$\; \buildrel < \over \sim \;$}
\newcommand{\gtsima} {$\; \buildrel > \over \sim \;$}
\newcommand{\lta} {\lower.5ex\hbox{\ltsima}}
\newcommand{\gta} {\lower.5ex\hbox{\gtsima}}

\def \ref {\noindent\hangindent=1.0in\hangafter=1}

\def\ltsima{$\; \buildrel < \over \sim \;$}
\def\simlt{\lower.5ex\hbox{\ltsima}} 
\def\gtsima{$\; \buildrel > \over \sim \;$}
\def\simgt{\lower.5ex\hbox{\gtsima}} 

\begin{document}
\title{High Energy Break and Reflection Features in the 
Seyfert Galaxy MCG+8-11-11}


\author{P. Grandi$^{1}$, F. Haardt$^{2}$,}

\author{G. Ghisellini$^{3}$, E. J. Grove$^{4}$, 
L. Maraschi$^{3}$, C. M. Urry$^{5}$}
\vskip2truecm

\noindent
$^{1}${Istituto di Astrofisica Spaziale, via E. Fermi 21, I-00044
Frascati (Roma) , Italy}\par
\noindent  
$^{2}${Dipartimento di Fisica, Universit\`a degli Studi di Milano, 
via Celoria 16, I-20133 Milano, Italy}\par
\noindent
$^{3}${Osservatorio Astronomico di Brera/Merate, via Brera 28, I-20121 
Milano, Italy}\par
\noindent
$^{4}${E. O. Hulburt Center for Space Research, Naval Research 
Laboratory, Washington D.C. 20375-5320, USA }\par
\noindent
$^{5}${STScI, 3700 San Martin Drive, Baltimore MD 21218, USA }\par
\noindent

\begin{abstract}
We present the results from ASCA and OSSE simultaneous observations of the
Seyfert 1.5 galaxy MCG+8-11-11 performed in August--September 1995. The ASCA
observations, split into two pointings of $\sim 10$ ksec each, indicate a
modest flux increase ($\sim 20\%$) in 3 days, possibly correlated to a
softening of the 0.6--9 keV spectrum. The spectrum obtained combining data from
both the observations is well described by a hard power law ($\Gamma \sim
1.64\pm 0.03$) absorbed by a column density slightly larger than the Galactic
value. An iron line at 6.4 keV of EW$\sim 400$ eV is also required by the data.
The simultaneous OSSE data are characterized by a much softer power law with
photon index $\Gamma=3.0^{+0.9}_{-0.8}$, strongly suggesting the presence of a
spectral break in the hard X/soft $\gamma-$ray band. In order to better
investigate the complex hard X-ray spectrum of this source, a total spectrum
has been produced by summing over all the observations of MCG+8-11-11 performed
by OSSE so far. A joint fit to the complete OSSE data sets and our ASCA data
clearly shows an exponential cut--off of the power law at $\sim 300$ keV ({\it
e}--folding energy), as well as a quite strong reflection component.
MCG+8-11-11 is then one of the few Seyfert galaxies where a spectral break
in the underlying continuum is detected {\it unambiguously}, fitting well in the
X--ray/$\gamma-$ray description of co--added spectra of Seyfert galaxies
(Zdziarski et al. 1995). Though the break could be due to downscattering of
X--rays on cold $e^+e^-$--pairs in a non--thermal plasma, the inferred low
compactness of this source favours instead thermal or quasi--thermal electron
Comptonization in a structured corona as the leading process of high energy
radiation production. 
 
\end{abstract}

{{\bf keywords}: galaxies: individual (MCG+8-11-11) --- galaxies: 
Seyfert -- X-rays: galaxies --- X-rays: general}


\section{Introduction}

MCG+8-11-11 is a nearby (z=0.021) face--on spiral galaxy classified as a
Seyfert galaxy of type 1.5. It is one of the brightest Seyfert galaxies in the
X-rays and shows flux variations by a factor 3--4 on time scale of years (Treves
et al. 1990, and references therein). It is one of the AGNs observed by HEAO1
for which Rothschild et al. (1983) suggested a canonical power law spectral
index $\Gamma\sim 1.7$ in the 2--165 keV band. EXOSAT observations of this
source (Treves et al. 1990) are well described by a simple power law (with
spectral slope ranging from 1.6 to 1.9), absorbed by column density with
Galactic value, $N_{\rm H}^{\rm Gal}$=2.03$\times10^{21}$ cm$^{-2}$ (Elvis,
Lockman, and Wilkes 1989).
EXOSAT revealed also a flux variation by a factor 2
from 1983 to 1986 (average unabsorbed flux F$_{\rm 1-8 keV}= 
6.4\times 10^{-11}$ ergs cm$^{-2}$ sec$^{-1}$) and a spectral softening with
increasing luminosity (Treves et al. 1990). 
Simultaneous observations performed by IUE showed no
correlation between the UV and X--ray emissions. 

Rosat observed this source only with the HRI instrument on 1991 March 19--20
for $\sim 20$ ksec. As we could not find any published result on this 
observation, we extracted and analyzed 
the data from the ROSAT public archive. 
No short time variability was detected during the pointing.
A spectral shape as suggested by the ASCA data in this paper 
($\Gamma=1.64$, $N_{\rm H}=2.5\times10^{21}$ cm$^{-2}$, see next \S 2.1) 
was used to convert the HRI count rate in energy flux units. 
We estimated an unabsorbed flux
$F_{\rm 1-8 keV}= 6.9 \times10^{-11}$ ergs cm$^{-2}$ sec$^{-1}$, 
comparable with the EXOSAT average flux level (Treves et al. 1990).

GINGA never pointed at MCG+8-11-11 so that no information on the
reflection component above 10 keV was available before summer 1995.
MCG+8-11-11 was detected in several balloon flights above 30 keV (Frontera et
al. 1979, Perotti et al. 1981, Ubertini et al. 1984, Perotti et al. 1990, 1997)
and is one of the four AGNs reported to show MeV emission (Perotti et al.
1981). The GRO satellite confirmed MCG+8-11-11 as an extremely intense AGN in
hard X--rays/soft $\gamma-$rays, with data extending up to $\sim 200$ keV.
Noticeably COMPTEL and EGRET failed to detected the source at MeV/GeV energies
(Lin et al. 1993, Maisack et al. 1993). Because of that, MCG+8-11-11 is
particularly suitable for checking the presence of an exponential cut--off, as
observed in the average high energy spectra of Seyfert galaxies 
(Madejski et al. 1995, Zdziarski et al. 1995, Gondek et al. 1996). 

The finding of an exponential cut--off in the soft $\gamma-$ray spectrum of
Seyfert galaxies has had a strong impact on the theoretical understanding of
the X-ray and $\gamma$-ray emission mechanisms, favoring thermal or
quasi-thermal Comptonization models (e.g. Liang 1979, 
Sunyaev \& Titarchuk 1980, Haardt \&
Maraschi 1991, 1993, Ghisellini, Haardt \& Fabian 1993, Titarchuk 1994,
Titarchuk \& Mastichiadis 1994, Svensson \& Zdziarski  1994, Haardt, Maraschi
\& Ghisellini 1994, Ghisellini, Haardt \& Svensson 1997). Zdziarski et al.
(1994) and Gondek et al. (1996) showed that also models with a mixture of
thermal and non--thermal electrons (and/or pairs) can describe the data
reasonably well. As pointed out by Ghisellini et al. (1993), the radiation
spectra provide information about the particle mean kinetic energy, rather than
about the actual particle distribution. Therefore a well posed question is
whether the mean Lorentz factor of cooling particles is small (say $\lta 10$)
or large ($\gta 100$), rather than whether particles are thermal or
non--thermal. Such a question can be answered by broad band X--ray
observations. We will then use the term "thermal" to indicate 
a particle distribution of small mean energy.

Here we report the results of ASCA and GRO observations of MCG+8-11-11 which
shows unambiguously a spectral break at $\sim 300$ keV. The implication
of this result on theoretical models will be discussed. 

\section{Observations and Data Analysis}

MCG+8-11-11 was observed continuously by OSSE from 1995 August 22 to September
7, and, within this period, by ASCA for $\sim 10$ ksec on September 3 and $10$ 
ksec on September 6. 

The reduction of ASCA data followed the standard procedures. For each
observation, a spectrum was accumulated using an extraction radius of 4' and 6'
for the SIS and GIS images, respectively. The background was estimated from
event files of source--free regions observed during the PV phase. 
Standard redistribution matrices (RMF; version 1994 November 9) 
were used for SIS and GIS. The ancillary files (ARF) have been produced using 
the program {\tt ascaarf} (version {\tt 2.64}).
Each spectrum was rebinned to a minimum of 25 counts per bin. 

OSSE data reduction also followed standard procedures. Energy spectra were 
accumulated in a sequence of two--minute measurements of the source field, 
alternated with two--minute offset pointing measurements of background fields. 
Prior to fitting, data from the four OSSE detectors were summed, grouped into 
12 keV--wide energy bins, and the appropriate response matrices were 
generated. 

All the SIS, GIS and OSSE data were fitted using the {\tt XSPEC} package,
version {\tt 10.0}. 
The ASCA spectral results reported in this paper refer to fits
simultaneously performed to data from all the 4 instruments. Each tested model
in {\tt XSPEC} was multiplied by a normalization constant in order to take into
account possible miscalibrations between the different instruments. SIS0 was
taken as the instrument with more reliable absolute flux, as suggested by the
latest ASCA calibration reports (http://heasarc.gsfc.nasa.gov/docs/asca). We
found a flux miscalibration less than 3$\%$ for the SIS1 and not larger than
10$\%$ for the GISs. 

If not otherwise indicated, the ASCA and OSSE quoted uncertainties correspond
to 90$\%$ confidence interval for one parameter of interest
($\Delta\chi^2=2.71$). 

\subsection {ASCA}

The X--ray flux of MCG+8-11-11 increased by $\sim 20\%$ between the two ASCA
observations. The count rates in the 0.6--9 keV range varied from
0.554$\pm0.007$ cts/s on September 3, to 0.669$\pm0.008$ cts/s on September 6
in SIS0. 

Both the observations are well described by an absorbed power law plus an iron
line from cold material (see Table~1 and Figure~1a and 1b). The average 
flux (unabsorbed flux,  F$_{\rm 1-8 keV}=2.6\times10^{-11}$ ergs 
cm$^{2}$ sec$^{-1}$) is a factor $\simeq 2.5$ lower than the ROSAT and 
EXOSAT flux levels. In both ASCA observations the value of $N_{\rm
H}$ is slightly larger than the Galactic column density $N^{\rm Gal}_{\rm
H}$=2.03$\times10^{21}$ cm$^{-2}$ (Elvis, Lockman $\&$ Wilkes 1989), 
even taking into account the systematic SIS overestimate of 
$N_{\rm H}=2-3 \times 10^{20}$ cm$^{-2}$. 
The residuals of September 3 data seem to 
suggest an absorption feature at $\sim 0.9-1$ keV.
Adding an edge to the model does not improve significatively 
the $\chi^2$ ($\chi^2=525$ for 506 d.o.f).
We therefore conclude that no significant spectral features other than the 
cold absorber are present in the soft X--ray spectrum of MCG+8-11-11.

The slightly different spectral slope between the two observations suggests a
steepening of the spectrum with increasing luminosity (see Table~1). However,
when fits to the data are performed in the 0.6--5.0 keV region (to avoid
continuum contamination from the 6.4 keV line), the spectral indices are
consistent within the uncertainties ($\Gamma_{1}=1.61\pm0.06$ and
$\Gamma_{2}=1.72\pm0.07$ on September 3 and 6, respectively). 
These are $90\%$  confidence statistic errors for two parameters of interest
($N_{\rm H}$ and $\Gamma$). The poor statistics due to short exposure times 
does not allow us to detect spectral variations at a high level of
significance.  

Finally, we combined data from the two observations. The ``total" SIS and GIS
spectra are characterized by an improved S/N, and therefore the continuum is
better constrained. An absorbed power law with a gaussian line is again a good
parameterization of the data. Adding a reflection component ({\tt PEXRAV} model
in {\tt XSPEC} from Magdziarz \& Zdziarski 1995) produces a significant
improvement of the $\chi^2$ statistics ($\Delta \chi^2 = 8$). However, both the
line parameters and the amount of reflection are poorly constrained (see
Table~1). 

\subsection {OSSE}

The OSSE spectrum is well described by a simple power law in the 50--200 keV
energy range ($\chi^2=4 $ for 10 d.o.f). The photon index is extremely steep
($\Gamma=2.97^{+0.91}_{-0.84}$) when compared to the ASCA slope, indicating the
presence of a break in the spectrum between 20 and 200 keV. The photon flux is
$F_{50-150 {\rm keV}} = 4.3\times10^{-4}$ photons cm$^{-2}$ sec$^{-1}$. 

In disagreement with the results from the HEAT balloon (Perotti et al. 1997),
no 112 keV line (interpreted as twice--scattered 511 keV annihilation line) was
detected, although the OSSE and HEAT observations are very close in time (HEAT
flew in July 1995), and the two data sets have similar continuum flux. 
OSSE upper limit for the 112 keV line was obtained assuming a power law
continuum and a gaussian line with a FWHM line width of 32 keV (as observed by
Perotti et al. 1997). The $95\%$ confidence upper limit is  $F_{112 {\rm
keV}}=9\times10^{-5}$ photons cm$^{-2}$ sec$^{-1}$, about a factor 10 below 
the HEAT value. As a further check, we also searched for a 170 keV feature in
emission (the once--scattered 511 keV line). As in the case of the lower energy
line, only upper limits could be derived ($F_{170 {\rm keV}}=7\times10^{-5}$
photons cm$^{-2}$ sec$^{-1}$). 

MCG+8-11-11 was detected by OSSE in two other epochs (viewing periods VP31 and
VP222, respectively) in 1992 June 11--25 and 1993 May 24--31 (Johnson 
et al. 1994). 
No significant luminosity variability appears between ours and previous 
observations (the $\chi^2$ probability that
the source is constant is $P_{\chi^2} \sim 0.1$). We then decided to
combine the data of each observing period to improve the S/N in the 50--200
keV range. The analysis of this ``total" OSSE spectrum indicates a spectral
shape in complete agreement with the result from the single observing period VP
427. As expected, the photon index is better constrained ($\Gamma_{\rm
tot}=2.70^{+0.42}_{-0.48}$). 

\subsection {ASCA plus OSSE Average Spectrum}  

We studied the overall spectral shape of MCG+8-11-11 by simultaneously fitting
the ``total'' ASCA spectrum (from the two 1995 September observations) and the
``total'' OSSE spectrum (from all the three observing periods). 

In order to investigate possible effects of flux miscalibration between ASCA
and OSSE, each model was tested assuming different values of the relative
SIS--OSSE normalization. No significant differences in the best fit parameters 
were found between the optimistic case of a perfect match between SIS and
OSSE fluxes (relative normalization = 1), and the worst examined case of
a miscalibration of $25\%$ (relative normalization = 0.75). For this reason, in
the following we will only discuss results obtained extrapolating the ASCA
models to the OSSE range (i.e., assuming relative normalization = 1). 

The best fit parameters are listed in Table~2. A simple power law plus a 6.4
keV emission line is a reasonable fit to the data ($\chi^2/$d.o.f$=915/871 $),
although a systematic deviation is evident in the model residuals to the OSSE
data (see Figure 2a). 

This result is not surprising since the $\chi^2$ test is not sensitive to the
(few) low statistics OSSE bins, and is essentially led by the higher quality
ASCA data. Adding  a reflection component from a face-on disk to the continuum
provides again an acceptable fit (see Table~2). The model residuals to the OSSE
data are better in this case, even if a further steepening of the 50--200 keV
spectrum is possibly required (see Figure 2b). 

We then checked whether an exponential cut--off power law rather than a simple
power law could represent the primary continuum. The $\chi^2$ significantly
improves as measured by an F--test ($\Delta \chi^2=16$ for one added parameter,
$P_{\rm F}>99.9$), and deviations of the OSSE data from the model are reduced
(Figure 2c). The {\it e}--folding energy is fairly constrained ($E_{\rm
cut}=266^{+90}_{-68}$ keV), as well as the amount of reflection as measured by
the ratio between the continuum and the reflection normalization ($R=
1.64^{+0.88}_{-0.78}$). 
The confidence region of the parameter $R$
indicates that a simple cut--off power law without reflection does not 
represent the OSSE data adequately .  

The energy-density spectrum of the SIS0 and OSSE data are plotted in Figure~3
with the best fit model to the data (cut--off power law plus reflection and iron
line). For comparison the simpler power law plus reflection and iron line model
(dotted line) is also shown. 

\subsubsection{Fit to a Comptonization Model}

As an exponential cut--off power law mimics the spectrum produced by hot
thermal electrons scattering off a low frequency radiation field (e.g., Sunyaev
\& Titarchuk 1980), we attempted to determine the electron parameters fitting
our data with a realistic Comptonization model. We then set up a Comptonization
model in plane parallel geometry, as outlined in Haardt (1993, 1994). 

The Compton spectrum produced in plane parallel geometry is a function of four
parameters, namely the vertical Thomson optical depth $\tau$, the electron
temperature $kT$, the soft photon input temperature $kT_{\rm BB}$ (assuming
black--body emission from the underlying accretion disk), and the viewing angle
$\mu\equiv \cos i$. The outgoing Comptonized spectrum is roughly a broken
power--law with an exponential cutoff at $E_{\rm cut}\sim 2kT$. The break in
the power--law part of the spectrum is due to the anisotropic nature of the
soft input photons. The break energy, $E_{\rm break}$, is just above the
average energy of twice--scattered photons, i.e. $E_{\rm break} \simeq
(16\Theta^2+4\Theta+1)^2 \times 2.7kT_{\rm BB}$ 
(Haardt 1993, Stern et al. 1995), which is typically much lower than 
the e-folding energy E$_{\rm cut}$ (we defined $\Theta\equiv kT/m_ec^2$). 
The slope of the power law below $E_{\rm break}$ depends on the viewing angle,
while it is steeper and almost angle--independent above. The effects of
anisotropic Compton scattering become more pronounced as $kT$ increases, and for
low inclination angles (Haardt 1993). 

The inclusion of the reflection hump, whose normalization depends also on the 
viewing angle and on the amplitude of the break in the power law, makes 
the model quite complicated to test.

Our data are not suited to allow  
a precise determination of all the model parameters 
because of  the limited statistics at high energies and the lack of data
between [10-50] keV, i.e. the region where the reflection peaks.
Indeed, we realized that different models computed for various inclination
angles and for various $kT_{\rm BB}$ are not distinguishable in term of 
$\chi^2$. In addition the amount of reflection, R, can vary 
from a minimum of about 1 up above 6, depending on $E_{\rm break}$. 
If the energy break occurs in the [10-50] keV, 
the reflection components has to be boosted in order to fill the gap above 
$E_{\rm break}$.  

Given these complications, we decided to fix some parameters. The black body
temperature was set to 5 or 20 eV, considering that larger values would produce
a soft excess not present in the data. The viewing angle was set to 30 or 60
degrees to test two possible inclinations of the accretion disk. Considering
that a possible contribution from an equatorial dust torus can not boost the
reflection more than a factor of $\sim 2$ (Ghisellini, Haardt \& Matt 1994;
Krolik, Madau \& Zycki 1994), we decided to freeze the normalization of the
reflection hump at $R=1.64$, according to the fit obtained with a cut--off
power law plus reflection. The data were then fitted searching for confidence
levels of the main parameters of the scattering cloud, namely $\tau$ and
$\Theta$. 

Results of our analysis are summarized in Table~3. Spectra obtained for
$kT_{\rm BB}=5$ eV and $\mu=0.5$ are close to a straight power law (times
exponential cut--off), while those obtained for $kT_{\rm BB}=20$ eV and
$\mu=0.867$ show the largest break in the power law due to significant
anisotropic Compton emission. The coronal temperature and optical depth are
tightly (anti)correlated, as expected, because their combination must produce a
spectrum with the observed slope. This is clearly displayed in Figure~4 where,
as an example, $\Theta-\tau$ confidence levels are plotted for $kT_{\rm BB}=20$
eV and $\mu=0.5$. Both the coronal temperature and the optical depth are well
constrained. Large values of $\Theta$ are not allowed as the spectrum would
show a strong break in the ASCA range, contrary to observations. Low $\Theta$
values are not permitted because they are in conflict with the spectral
cut--off observed in the OSSE data. It should be noted that the upper limit on
the temperature depends on the assumed viewing angle and black body
temperature. Table~3 shows that $\Theta$ best fit value is shifted to higher
values when $kT_{\rm BB} = 5$ eV with respect to models computed for $kT_{\rm
BB}=20$ eV. On the contrary, the lower limit ($\Theta \gta 0.2$) does not
depend on the assumed input parameters. Indeed, when two parameters of interest
are considered, the 99\% confidence lower limit for $\Theta$ is $\simeq 0.2$
also in cases of $kT_{\rm BB}=5$ eV. 

\section{Discussion}

MCG+8-11-11 is one of the few Seyfert galaxies where a spectral break is
clearly detected in the hard X-ray spectrum. The best fit to the ASCA and OSSE
data is an absorbed power law ($\Gamma=1.73\pm0.06$) with exponential cut--off
($E_{\rm cut}=266^{+90}_{-68}$ keV), plus a reflection component
($R=1.64^{+0.88}_{-0.78}$) and a cold iron line (EW=230$^{+222}_{-99}$ eV).

The ASCA+OSSE observations can be readily interpreted in the framework of
thermal Comptonization. A detailed Comptonization model in plane parallel
geometry gives a temperature $kT\simeq 150\pm 50$ ($250 \pm 50$) keV and an
optical depth $\tau \simeq 0.3^{+0.2}_{-0.1}$ ($0.15\pm 0.05$) for $kT_{\rm
BB}=20$ (5) eV. These values correspond to a Compton parameter $y\simeq 0.9$,
larger than allowed in a model where Comptonization occurs in a homogeneous
corona above an accretion disk (Haardt \& Maraschi 1991). In fact, in this
case, the energy balance requirements yeald $y\simeq 0.6$. We therefore
conclude that a photon starved source (i.e., a source with reduced feedback of
the disk emission on the Comptonizing region) is required to obtain a spectrum
as flat as observed. A structured corona (Galeev, Rosner \& Vaiana 1979, Haardt
et al. 1994) must be invoked, and indeed the Compton parameter we have derived
from the data is basically consistent with what obtained assuming hemispherical
sources located above an accretion disk (Stern et al. 1995, and Poutanen \&
Svensson 1996). Alternative geometries, probably appropriate for Galactic
sources (e.g., Gierlinski et al. 1997 and Dove et al. 1997 for the hard state
of Cygnus X--1), would have problems in explaining the prominent reflection
features observed in MCG+8-11-11 and the strong iron line (see next). 

The normalizations of the two cold matter signatures (hump and iron line) are
suggestive of extra reflection other than from the accretion disk, though the
more comfortable $\Omega/2\pi=1$ value is within the 90\% confidence level for
both parameters. The extra reflection flux (if real) may have several origins:
it may arise from reflection of X--rays in an extended molecular torus
(Ghisellini et al. 1994; Krolik et al. 1994), and/or it may be (partly) due to
anisotropic Compton effect (Haardt 1993). Due to the latter effect, an
equivalent width of the Fe K line of $\simeq 200$ eV is expected for a coronal
temperature of $\simeq 250$ keV, assuming an average soft photon temperature of
$\simeq 50$ eV (Poutanen, Svensson \& Stern 1997). 

The significance of spectral variability between the two ASCA observations is
marginal. Nevertheless it is interesting to make some speculation. First, we
note that the possible ``steeper when brighter" behaviour is the typical trend
of Seyfert galaxies in the ASCA band (e.g. Treves et al. 1990, Leighly et al.
1996). Second, the best fit value $\Delta \Gamma\simeq 0.1$, coupled with a
variation of 20\% of the 2--10 keV flux, is consistent with variations of $\tau$
at constant luminosity (see Figure 7 of Haardt et al. 1997). Such 
interpretation is also supported by long term variability results. 
In fact, while the average 1--8 keV flux of the ASCA observations 
was a factor $\simeq 2.5$ lower than the flux detected by ROSAT, 
the soft $\gamma-$ray flux did not show apparent variations 
between the three OSSE pointings. This could indicate that 
long term flux variations are larger in the low energy part of the 
X--ray spectrum, i.e. the spectrum pivots at high energy. 
Since $\Gamma<2$, the luminosity of the source would be then roughly 
constant. We note that 
similar variability features were also present  
in the EXOSAT data (Treves et al. 1990), showing a pivoting of the 
(extrapolated) power law at $\simeq 50$ keV.

The 20\% flux variation between the two ASCA observations, if linearly
extrapolated, leads to a doubling timescale $\Delta t_2\simeq 15$ days,
consistent with the earlier results by Done \& Fabian (1989), giving a
compactness value of $\ell \sim 0.1$, where $\ell\equiv {\sigma_{\rm T} \over
m_{\rm e}c^3}L/c\Delta t_2$. On the other hand, assuming a pair 
dominated plasma, we can translate the maximum
coronal temperature allowed by the data (i.e. $kT\simeq 250$ keV) into the
minimum possible value of the compactness, namely $\ell \gta 50$
for a hemispherical source (see Figure 3 of Svensson 1996). This may indicate
that pair production is not important in MCG+8-11-11. 

The unambiguous detection of a break in the high energy spectrum of MCG+8-11-11
is suggestive of a thermal origin of the intense X--ray emission in this
source. Still, a spectral break could also be consistent with non thermal pair
models, as, e.g., the case of IC4329A described by Zdziarski et al. (1994), and
in general OSSE data alone are not sufficient to discriminate between the two
models. However, in order to produce the observed break in the hard X--ray
spectrum, non--thermal models require a large value of $\ell$, not supported by
variability results. We conclude that, at least in this slowly variable source,
thermal models are favoured. 

A criticism to the previous conclusion can arise from the HEAT observation by
Perotti et al. (1997). In fact the claimed detection of an emission line at
$\simeq 112$ keV (interpreted as due to double Compton scattering of an unseen
511 keV annihilation line) may be due to a non thermal origin of the X--ray
emission in MCG+8-11-11. The HEAT line photon flux is $9\times 10^{-4}$ photons
cm$^{-2}$ s$^{-1}$, while in the OSSE data any similar line has a 95\%
confidence upper limit flux of $0.9 \times 10^{-4}$ photons cm$^{-2}$ s$^{-1}$.
The continuum in the OSSE data is almost at the same level as in the HEAT
observation. This implies that the line, if real, is variable. However, as
Perotti et al. (1997) point out, there is also the possibility that the line
comes from an outburst of a galactic source in the HEAT field of view, reaching
flux levels comparable to those of 1E 1740.7--2942 and Nova Muscae. 

In conclusion, MCG+8-11-11 shows the typical broad band X--ray/$\gamma-$ray
spectrum of Seyfert 1 galaxies. The presence of a strong reflection component
indicates that a large fraction of the primary X--rays are reprocessed by cold
matter, possibly the nearby accretion disk. The spectral break at $\simeq 270$
keV is consistent with cooling of an active structured corona via thermal
Comptonization. Large compactness of the corona is not supported by the data,
on the basis of the weak variability in X--rays on short time scales. Therefore
pairs should be only a marginal component of the X--ray emitting plasma.

\acknowledgments
We gratefully thank S. Molendi for performing the ROSAT HRI data reduction, and
the anonymous referee for useful suggestions. PG and CMU acknowledge support
from NASA grants NAG5-2510 and NAG8-1037.

\newpage

\figcaption{The ASCA SIS spectrum (upper panel) and the 
residuals (lower panel) when an absorbed power law plus iron line model 
are fitted to the data of September 3 (a) and September 6 (b).}

\figcaption{The ratio of the OSSE residuals for a) a simple power law
plus gaussian model, b) a power law plus reflection and gaussian line model, c)
a cut--off power law plus reflection and gaussian line model.}

\figcaption{Energy density unfolded spectrum of the combined SIS0 and OSSE
data. The best fit to the data is a cut--off power law plus reflection and
gaussian line (solid line). A simpler power law plus reflection and gaussian
line model (dotted line) is also shown for comparison.}

\figcaption{$\Theta$--$\tau$ contour plot (68\%, 90\% and 99\% 
confidence levels for two parameters of interest) 
for Comptonization model with $kT_{\rm BB}=20$ eV and $\mu=0.5$.}

\newpage

\setlength{\textwidth}{14truecm}
\setlength{\textheight}{19truecm}
\setlength{\topmargin}{-3.2truecm}
\setlength{\oddsidemargin}{-1.0truecm}
\centering
\tiny
\begin{table}
 \begin{tabular}{lcccccccc}
 \multicolumn{9}{c}{{\Large {\bf Table 1:} ASCA Spectral Fit Parameters}} \\
 &&&&&&&&\\ \hline\hline
  \multicolumn{1}{c}{Date}
 &\multicolumn{1}{c}{Model$^a$}
 &\multicolumn{1}{c}{$\Gamma$}
 &\multicolumn{1}{c}{$N_{\rm H}$}
 &\multicolumn{1}{c}{$E_{\rm Fe}$}
 &\multicolumn{1}{c}{$\sigma_{\rm Fe}$}
 &\multicolumn{1}{c}{EW}
 &\multicolumn{1}{c}{R$^b$}
 &\multicolumn{1}{c}{$\chi^2$/(d.o.f)}\\
 &&&&&&&&\\
 &
 &
 &$10^{21}$ cm$^{-2}$
 &\multicolumn{1}{c}{keV}
 &\multicolumn{1}{c}{keV}
 &\multicolumn{1}{c}{eV}
 &
 &\\
  \hline
 &&&&&&&&\\
 
 Sept 3 & PL+GA & 1.57$\pm0.04$ & 2.32$^{+0.25}_{-0.23}$& 6.37$^{+0.09}_{-0.06}$&
 0.15$^{+0.16}_{-0.07}$ & 392$^{+129}_{-108}$& & 526/508\\
 &&&&&&&&\\
 Sept 6 & PL+GA  & 1.69$^{+0.06}_{-0.04}$ & 2.65$^{+0.19}_{-0.12}$
 &6.42$^{+0.15}_{-0.12}$ & 0.31$^{+0.69}_{-0.13}$ & 401$^{+171}_{-149}$ 
 & & 594/558\\
 &&&&&&&&\\
 Sept 3+6& PL+GA & 1.64$\pm0.03$ & 2.50$^{+0.16}_{-0.15}$ &
 6.43$\pm0.08$ & 0.27$^{+0.10}_{-0.09}$ & 380$^{+102}_{-83}$ & & 896/860\\
 &&&&&&&&\\
 Sept 3+6 & PL+REF+GA &1.74$^{+0.08}_{-0.03}$& 2.76$^{+0.00}_{-0.04}$& 
 6.39$^{+0.10}_{-0.15}$& 
 0.20$\pm0.20$ &232$\pm117$  & 1.56$^{+1.54}_{-0.90}$ & 888/859\\
 &&&&&&&&\\
 \hline
 &&&&&&&&\\
 \multicolumn{9}{l}{$^a$-- PL = Power Law, GA = Gaussian Profile, 
 REF = Reflection (PEXRAV model with E$_{\rm cut}=0$).}\\
&&&&&&&&\\
 \multicolumn{9}{l}{$^b$-- R is the ratio between the continuum and the 
 reflection normalization.}\\ 

 \end{tabular}
\end{table}

\newpage
\setlength{\textwidth}{16truecm}
\setlength{\textheight}{19truecm}
\setlength{\topmargin}{-3.2truecm}
\setlength{\oddsidemargin}{-0.8truecm}
\centering
\begin{table}
 \begin{tabular}{lcccccccc}
 \multicolumn{9}{c}{{\Large {\bf Table 2}: 
ASCA + OSSE: Spectral Fit Parameters}}\\
 &&&&&&&&\\
 &&&&&&&&\\ \hline\hline
 \multicolumn{1}{c}{Model$^a$}
 &\multicolumn{1}{c}{$\Gamma$}
 &\multicolumn{1}{c}{$N_{\rm H}$}
 &\multicolumn{1}{c}{$E_{\rm Fe}$}
 &\multicolumn{1}{c}{$\sigma_{\rm Fe}$}
 &\multicolumn{1}{c}{EW}
 &\multicolumn{1}{c}{R$^b$}
 &\multicolumn{1}{c}{E$_{\rm cut}$}
 &\multicolumn{1}{c}{$\chi^2$/(d.o.f)}\\
 &&&&&&&\\
 &
 &$10^{21}$ cm$^{-2}$
 &\multicolumn{1}{c}{keV}
 &\multicolumn{1}{c}{keV}
 &\multicolumn{1}{c}{eV}
 &
 &
 \\
 \hline
&&&&&&&&\\
 PL + GA& 1.65$^{+0.02}_{-0.03}$& 2.53$^{+0.15}_{-0.13}$& 6.43$\pm0.08$&
 0.28$^{+0.10}_{-0.21}$&470$^{+86}_{-97}$& & & 915/871\\
 &&&&&&&&\\
 PL + GA + REF &1.71$^{+0.03}_{-0.07}$ &2.73$^{+0.18}_{-0.00}$  &6.4(f)  & 0.3(f)
  &397$^{+81}_{-73}$   &0.50$^{+0.35}_{-0.50}$ & & 913/872\\
 &&&&&&&&\\
 CUTPL + GA  & 1.73$\pm0.06$& 2.75$\pm0.22$ & 6.4(f) & 0.3(f)
 & 230$^{+222}_{-99}$  & 1.64$^{+0.88}_{-0.78}$ &266$^{+90}_{-68}$ & 897/871\\
 &&&&&&&&\\
 \hline
 &&&&&&&&\\
 \multicolumn{9}{l}{$^a$-- PL = Power Law, GA = Gaussian Profile, 
 REF = Reflection (PEXRAV model with E$_{\it cutoff}=0$. )}\\
 &&&&&&&&\\
 \multicolumn{9}{l}{CUTPL = PEXRAV model with E$_{\rm cut}$ free to vary.}\\
 &&&&&&&&\\
 \multicolumn{9}{l}{$^b$-- R is the ratio between the continuum and the 
 reflection normalization.}\\
 \end{tabular}
\end{table}

\newpage
\setlength{\textwidth}{16truecm}
\setlength{\textheight}{19truecm}
\setlength{\topmargin}{-3.2truecm}
\setlength{\oddsidemargin}{-0.8truecm}
\centering
\begin{table}
 \begin{tabular}{cccccccc}
 \multicolumn{8}{c}{{\Large {\bf Table 3}: 
ASCA + OSSE: Compton Model Fit Parameters}}\\
 &&&&&&&\\
 &&&&&&&\\ \hline\hline
 \multicolumn{2}{l}{Input Parameters}
 &\multicolumn{1}{c}{$N_{\rm H}$}
 &\multicolumn{1}{c}{EW$^a$}
 &\multicolumn{1}{c}{$\Theta$}
 &\multicolumn{1}{c}{$\tau^{b}$}
 &\multicolumn{1}{c}{R$^c$}
 &\multicolumn{1}{c}{$\chi^2$/(d.o.f)$^d$}\\
 \multicolumn{1}{c}{$kT_{\rm BB}$}
 &\multicolumn{1}{c}{$\mu$}
 &&&&&&\\
 \multicolumn{1}{c} {eV}
 &&\multicolumn{1}{c} {$10^{21}$ cm$^{-2}$}
 &\multicolumn{1}{c}{eV}
 &&&&
 \\
 \hline
 &&&&&&&\\
5&
0.5 & 
2.85$^{+0.17}_{-0.18}$ &
306$^{+134}_{-144}$ &
0.53$^{+0.09}_{-0.10}$ &
0.12$^{+0.09}_{-0.02\star}$ &
1.64(f)&
167/190\\
&&&&&&&\\
5 & 
0.867 & 
2.53$^{+0.21}_{-0.18}$ &
324$^{+132}_{-146}$ &
0.50$^{+0.07}_{-0.06}$ &
0.14$^{+0.08}_{-0.02}$ &
1.64(f)&
167/190\\
&&&&&&&\\
20 & 
0.5 & 
2.78$^{+0.20}_{-0.19}$ &
314$^{+115}_{-149}$ &
0.29$^{+0.14}_{-0.04}$ &
0.40$^{+0.12}_{-0.22}$ &
1.64(f)&
166/190\\
&&&&&&&\\
20 & 
0.867 & 
2.52$^{+0.18}_{-0.17}$ &
303$^{+130}_{-133}$ &
0.32$^{+0.05}_{-0.14}$ &
0.32$^{+0.20}_{-0.10}$ &
1.64(f)&
167/190\\
&&&&&&\\
 \hline
 &&&&&&\\
 \multicolumn{8}{l}{$^a$-- $E_{\rm Fe}=6.4$(f) keV, 
$\sigma_{\rm Fe}=0.3$(f) kev.}\\
 &&&&&&\\
 \multicolumn{8}{l}{$^b$-- $\star$ 
  indicates values pegged to model limits.}\\
 &&&&&&\\
 \multicolumn{8}{l}{$^c$-- R is the ratio between the continuum and the 
 reflection normalization.}\\
 &&&&&&\\
 \multicolumn{8}{l}{$^d$-- CUTPL+GA gives $\chi^2/$(d.o.f.)=166/189.}\\
 \end{tabular}
\end{table}

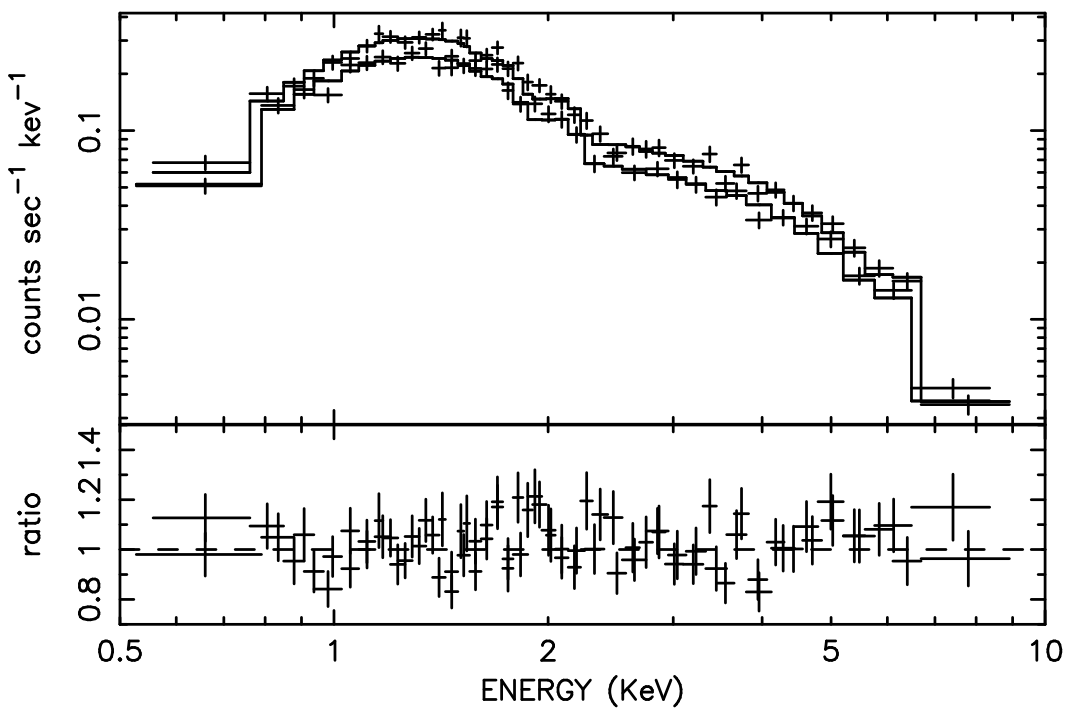
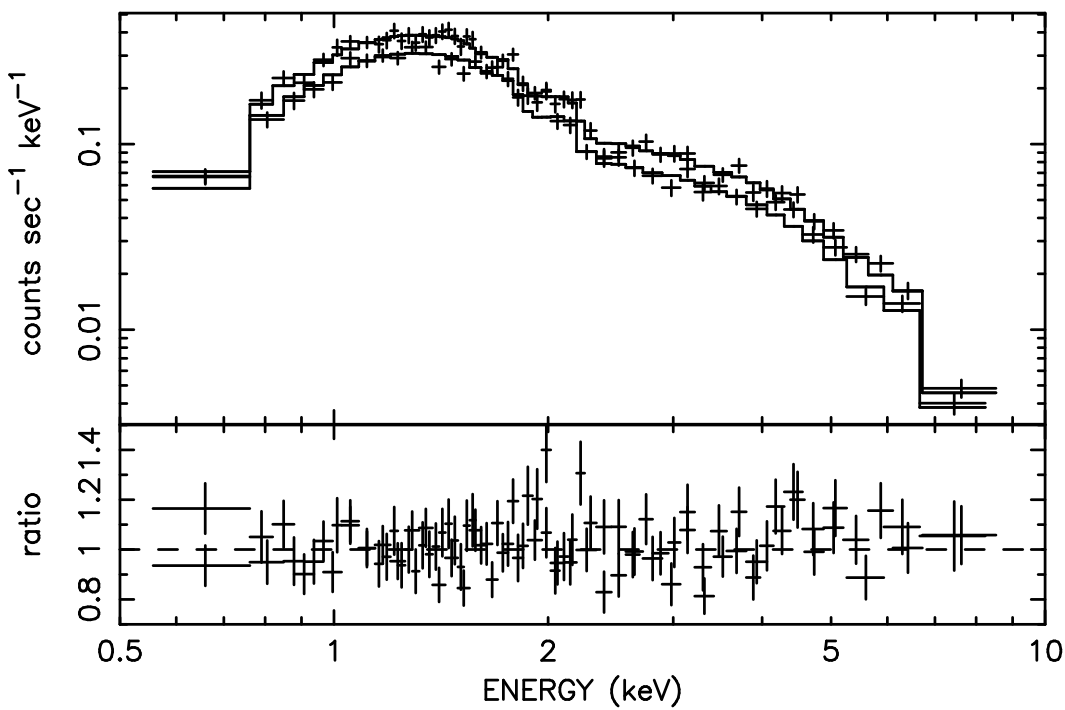
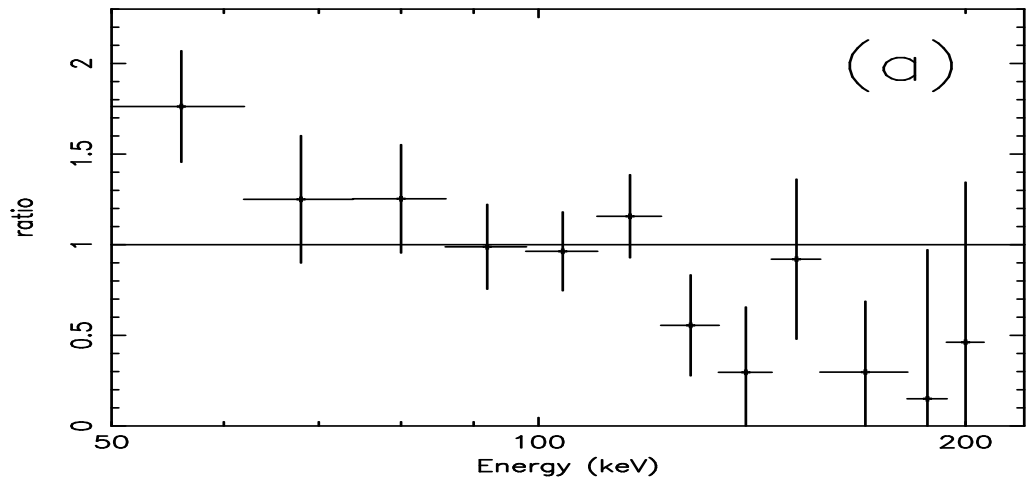
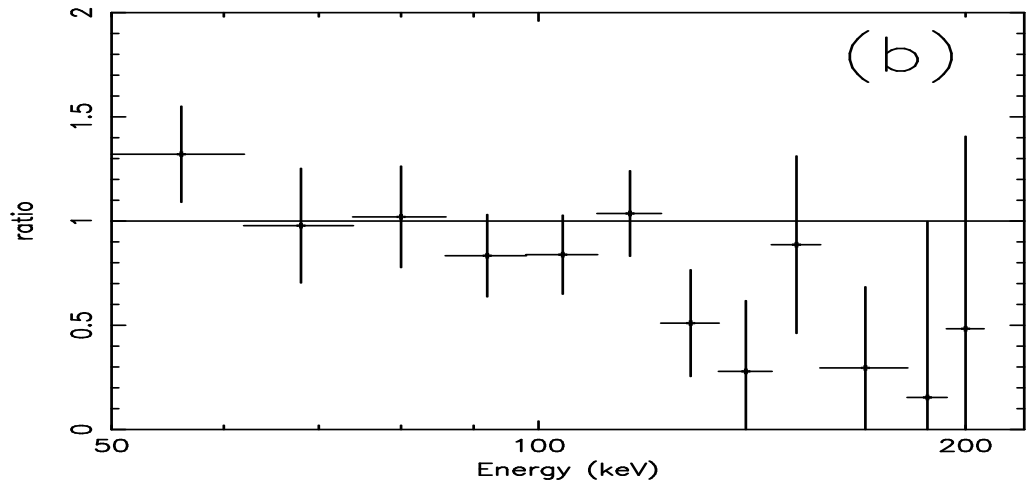
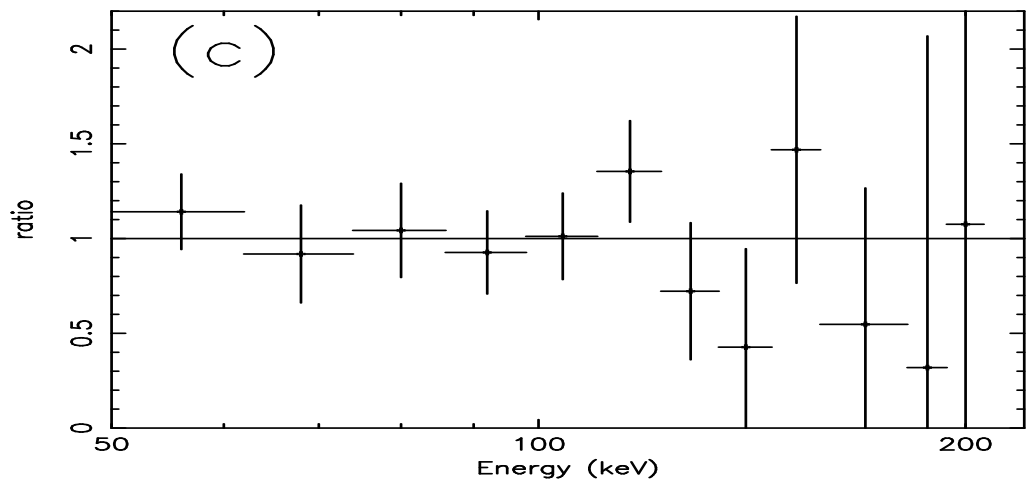
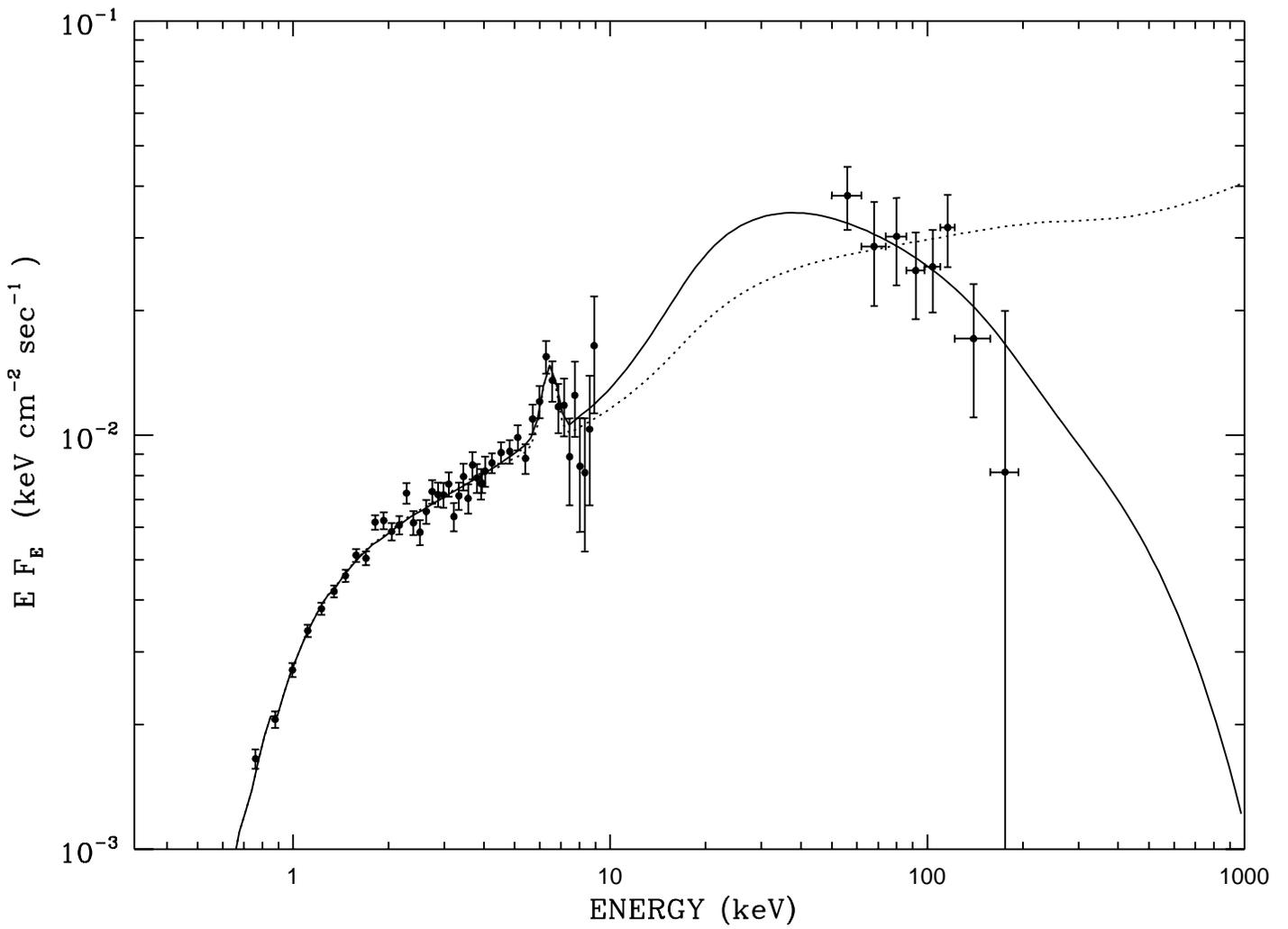
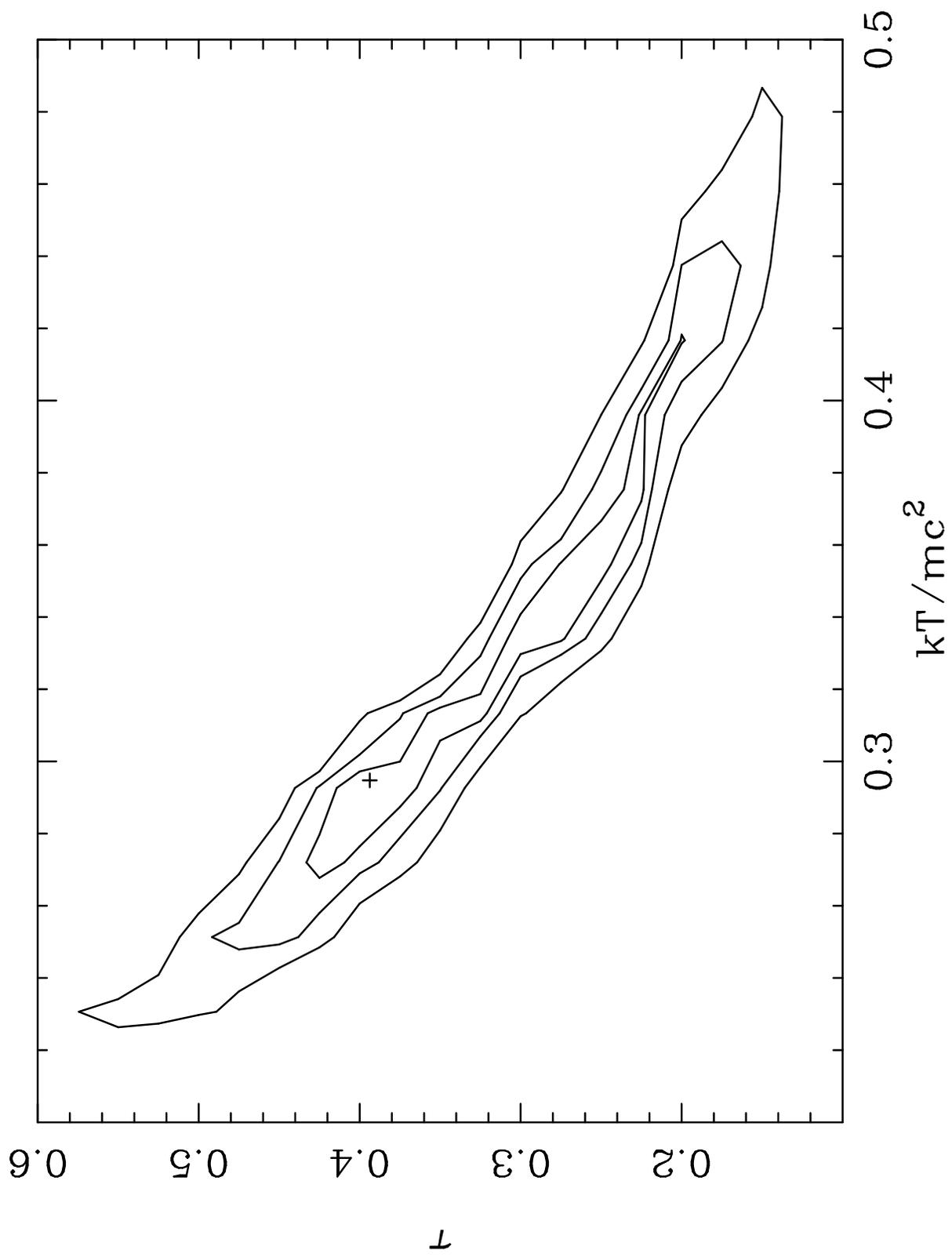
\end{document}